\documentclass[
               twocolumn,
               noshowpacs,          
               nopreprintnumbers,     
               aps,                 
               prd,                 
               a4paper,             
               superscriptaddress,  
               nofootinbib,         
               tightenlines,        
               floats,floatfix      
               ]{revtex4}

\usepackage{amsmath, amsfonts, amsthm, amssymb, graphicx}
\usepackage{epstopdf}
 \usepackage{slashed}
 \usepackage{graphicx,epsfig}
\usepackage{amsmath}
\usepackage {amssymb}
\usepackage[utf8]{inputenc}
\usepackage{hyperref}
\usepackage[francais]{babel}
\usepackage[T1]{fontenc}

\newmuskip\pFqmuskip

\newcommand*\pFq[6][8]{%
  \begingroup 
  \pFqmuskip=#1mu\relax
  \mathchardef\normalcomma=\mathcode`,
  \mathcode`\,=\string"8000
  \begingroup\lccode`\~=`\,
  \lowercase{\endgroup\let~}\pFqcomma
  {}_{#2}F_{#3}{\left[\genfrac..{0pt}{}{#4}{#5};#6\right]}%
  \endgroup
}
\newcommand{\pFqcomma}{{\normalcomma}\mskip\pFqmuskip}

\newcommand{\be}{\begin{eqnarray}}
\newcommand{\ee}{\end{eqnarray}}
\newcommand{\bea}{\begin{eqnarray}}
\newcommand{\eea}{\end{eqnarray}}

\def\n{\nu}

\begin{document}

\title{Jackiw-Teitelboim Gravity as a Noncritical String}

\author{Eoin Dowd}
\affiliation{Department of Physics, New York University, 726 Broadway, New York, NY10003, USA.}

\author{Gaston Giribet}
\affiliation{Department of Physics, New York University, 726 Broadway, New York, NY10003, USA.}



\begin{abstract}
Jackiw Teitelboim (JT) gravity has proven to be an excellent tool for investigating aspects of quantum gravity and black hole physics. In recent years, the study of JT gravity and its deformations has helped us learn about the different contributions of geometries in the gravitational path integral, the quantum gravity Hilbert space, the space-time factorization problem, the role of averaging in holography, the black hole information paradox, and the matrix models. All this motivates the exploration of the JT gravity in different setups, with and without matter. Here, we consider JT gravity conformally coupled to Liouville field theory and matter fields. This model admits to be interpreted as a non-critical string theory on a linear dilaton background with a tachyonic Liouville potential along a null direction. The constant curvature constraint of JT gravity results in a neutralization of the Liouville mode, which makes it possible to compute the four-point correlation function of the theory analytically. Here we give the explicit derivation of the four-point function and briefly comment on its properties, such as monodromy invariance, crossing symmetry, factorization, and limits.   
\end{abstract}

\maketitle

\section{Introduction}

Jackiw-Teitelboim (JT) gravity \cite{JT1, JT2, JT3} has garnered renewed attention in recent years due to its connections to various areas of theoretical physics. Originally proposed as a two-dimensional model of gravity coupled to a dilaton field, JT gravity serves as a simplified yet powerful tool for studying fundamental concepts in quantum gravity, black hole physics, condensed matter, quantum information, and holography; see \cite{duality0}-\cite{duality16} and references therein and thereof; for a recent review, see \cite{review}. 

The model features a gravitational action that, despite its simplicity, encapsulates rich physics. Its mathematical tractability allows for exact analytical treatments, making it an ideal testing ground for exploring semi-classical and quantum gravity, especially through its connection to the Sachdev-Ye-Kitaev model \cite{SY,K,Kitaev}. In recent years, the study of JT gravity has allowed us to expand our knowledge about the path integral formulation of semi-classical gravity, the Hilbert space of quantum gravity, the causal structure of the spacetime within the context of holography, black hole thermodynamics, and matrix models. All this motivates the exploration of JT gravity and its extensions further. In this paper, we will study JT gravity conformally coupled to Liouville field theory and additional matter fields. This model admits to be interpreted as a non-critical string theory on a linear dilaton background with a tachyonic Liouville potential along a null direction \cite{Mazzitelli, Gaston}. The constant curvature constraint of JT gravity results in a neutralization of the Liouville mode, which makes it possible to compute the four-point correlation function of the theory analytically. Here we give the explicit derivation of the four-point function and discuss its properties, such as monodromy invariance, crossing symmetry, factorization, and limits.   

The paper is organized as follows: In section II, we will introduce the theory we will be concerned with, which mainly consists of JT gravity interacting with Liouville field theory through a Weyl coupling of the metric. In section III, we will review the formulation of the 2D gravity theory as a non-critical string $\sigma $-model in a non-trivial dilaton-tachyon background. The background exhibits a light-like Liouville direction which only interacts with transversal excitations. This simplifies the problem of computing correlation functions enormously, as we comment in section IV. The explicit computation of the correlation functions is given in section V, where analytic formulae for the three- and four-point functions are presented. Section VI contains some final remarks.

\section{Two-dimensional Gravity}

We will consider a theory consisting of JT gravity
coupled to Liouville field theory and additional matter fields. The action of the full theory is
\begin{equation}
S=S_{JT}+S_{L}+S_{\text{matter}}+S_{\text{ghost}} \, , \label{sjt}
\end{equation}
with the JT action \cite{JT1, JT2}
\begin{eqnarray}
S_{JT} =\frac{1}{2\pi }\int_{M}d^{2}x\sqrt{\hat{g}}\,\varphi\, ( \hat{R}+%
{\Lambda}) +\frac{\theta }{4\pi }\int_{M}d^{2}x\sqrt{\hat{g}}%
\,\hat{R}\label{SJT}
\end{eqnarray}
coupled to the Liouville action \cite{Liouville}
\begin{eqnarray}
S_{L} &=&\frac{1}{2\pi }\int_{M}d^{2}x\sqrt{g} \,g^{\alpha \beta
}\partial _{\alpha }\phi \partial _{\beta }\phi + \frac{Q}{2\pi }\int_{M}d^{2}x\sqrt{g}\,R\,\phi\, \nonumber \\
&& +\,{\mu }
\int_{M}d^{2}x\sqrt{g}\,e^{\gamma \phi }\label{SL}
\end{eqnarray}
together with matter fields
\begin{eqnarray}
S_{\text{matter}} =\frac{1}{4\pi }\int_{M}d^{2}x\sqrt{g}\,g^{\alpha \beta }\partial
_{\alpha }X^{a}\partial _{\beta }X^{b}\delta _{ab}\, ,
\end{eqnarray}
where $\alpha , \beta =0,1$ and $a,b=2,3,...,D-1$. Matter content is given by $D-2$ space-like scalar fields $X^{a}$. The coupling constants $\theta $, $\mu $, $Q$, $\gamma $ are real, with $\gamma\neq -1$. Relations among some of these couplings are required for the theory to exhibit special properties, such as conformal invariance and self-duality \cite{Gaston}.

We will consider the conformal theory on a closed manifold $M$. The term in the action with coupling $\theta$ gives the Euler characteristic of $M$, denoted $\chi (M)$. The metrics in (\ref{SJT}) and (\ref{SL}) are related by a Weyl transformation
\begin{equation}
\hat{g}_{\alpha \beta }=e^{\gamma \phi }g_{\alpha \beta }  \label{gauge}
\end{equation}
which involves the Liouville field $\phi $. The field $\varphi $ is the JT scalar, which enters in $S$ as a Lagrange multiplier that yields the constant curvature constraint
\begin{equation}
\hat{R}+{\Lambda}=0,\label{vcc}
\end{equation}
together with a second-order non-linear field equation for $\varphi $. Shifting the zero mode of $\varphi $ changes the value of $\theta $; shifting the zero mode of $\phi $ produces a similar effect but at the price of rescaling $\mu $. 

The ghost contribution to the action is also required and can be
expressed in terms of the $b$-$c$ system as 
\begin{equation}
S_{\text{ghost}}=\frac{1}{2\pi }\int_{M}d^{2}x\sqrt{g}\,g^{\alpha \beta }c^{\rho
}\nabla _{\alpha }b_{\beta \rho }\,.
\end{equation}
Hereafter we will omit the ghost contribution for brevity.

On $M$, we use coordinates $x^{\alpha}$ with $x^0=t$ and $x^1=x$; as usual, $\partial_{\alpha }$ stands for the derivative with respect to $x^{\alpha}$. We are going to consider the theory on the Riemann projective sphere, $\mathbb{CP}^ 1$, choosing the metric $g_{\alpha \beta}$ to be the locally flat metric written in the standard complex variables $z=x^0+ix^1$, $\bar z= x^0-ix^1$, with $d^2z=\frac i2 dzd\bar z$, and $\partial = \frac{\partial}{\partial z}$, $\Bar{\partial} = \frac{\partial}{\partial \bar z}$. We thus have $\chi(M)=2$.

\section{Non-Critical String}

The theory defined by (\ref{sjt}) can be written as a string theory $\sigma $-model on a non-trivial background. To see this, let us redefine fields as follows
\begin{eqnarray}
\phi =\frac{ X^{1}+X^{0}}{ \sqrt{2(1+\gamma )}} \, , \ \ \ \ 
\varphi =\frac{ X^{1}-\xi X^{0}}{ \sqrt{2(1+\gamma )}} \label{var2}
\end{eqnarray}
with $\xi =1+{2}{\gamma }^{-1}$. These new variables are well-defined in terms of the original coordinates $\varphi $ and $\phi $ provided $\gamma \neq -1$. In terms of the new target space coordinates, the action above takes the form
\begin{equation}
S=S_{0}+S_{I}+\theta \chi (M)+S_{\text{ghost}}  \label{sx}
\end{equation}
with the Gaussian contribution
\begin{eqnarray}
S_{0} &=&\frac{1}{4\pi }\int_M d^{2}z\left( \eta _{\mu \nu }\partial X^{\mu }%
\bar{\partial}X^{\nu }+RQ_{0}X^{0}+RQ_{1}X^{1}\right) ,  \nonumber
\end{eqnarray}
and the interaction term
\begin{eqnarray}
S_{I} =\frac{1}{4\pi }\int_M d^{2}z\left( \hat\Lambda
X^{1}-\xi \hat\Lambda X^{0}+4\pi\mu \right) e^{2b (X^{0}+X^{1})} \,\, ,\nonumber
\end{eqnarray}
where $\eta _{\mu \nu }$ denotes the mostly-plus, $D$-dimensional Minkowski metric ($\mu , \nu =0,1,2,..., D-1$). Also, we have rescaled the cosmological constant as
\begin{equation}
\hat\Lambda =\sqrt{\frac{2}{1+\gamma }} \Lambda \, ,
\end{equation}
and introduced the following notation 
\begin{eqnarray}
Q_{0} &=& \sqrt{\frac{2}{1+\gamma }} (Q-1-2\gamma^{-1})\, ,
\\
Q_{1} &=&\sqrt{\frac{2}{1+\gamma }} (Q+1)\, .
\end{eqnarray}
As said, conformal invariance demands additional constraints among the couplings. We write
\begin{equation}
2b=\frac{\gamma}{\sqrt{2(1+\gamma)}}\, .\label{ertyui}
\end{equation}

The field equations associated with the fields $X^0$ and $X^1$ can be combined to reproduce the constant curvature constraint of JT theory. More precisely, one obtains the Liouville equation
\begin{equation}
2b\, \partial\bar\partial X^+ = R +\Lambda \, e^ {2bX^+}\, ,
\end{equation}
$X^{+}=X^0+ X^1$, which written in terms of (\ref{var2}) and using (\ref{ertyui}) agrees with (\ref{vcc}). To show this be have to be reminded of (\ref{gauge}) implying 
\begin{equation}
\hat{R}=e^{-\gamma \phi } (R-\gamma \nabla^ 2 \phi ).
\end{equation}

Here, we will consider the theory with $\Lambda =0$. For the case $\Lambda \neq 0 $ see reference \cite{Gaston}. For $\Lambda =0$, the action above can interpreted as a string worldsheet $\sigma$-model in the $D$-dimensional dilaton-tachyon background
\begin{equation}
\Phi(X)= \frac{1}{2}Q_{\mu}X^{\mu} +\frac{\theta}{2}\, , \ \ \ \ \ T(X)= {\pi\mu}  \, e^{2b_{\nu}X^{\nu}}
\end{equation}
with $Q_{\mu}=Q_0\,\delta_{\mu}^0+Q_1\,\delta_{\mu}^1$, $b_{\nu}=b\,(\delta_{\nu}^0+\delta_{\nu}^1)$ with $\mu , \nu =0,1,2,...,D-1$ and $\alpha ' =2$. This describes a flat, linear dilaton configuration with a Liouville type potential along the light-like direction $X^+$; see Figure 1. The tachyon potential in the case $\Lambda \neq 0$ has been studied in \cite{Gaston}, where it was shown how the condition for $T(X)$ to be a puncture primary operator that follows from the operator product expansion with the stress tensor actually coincides with the vanishing of the $\beta $-function at 1-loop. Hereafter, we take $\Lambda =0$.

The central charge of the worldsheet theory, excluding the ghost contribution, is
\begin{equation}
c\,=\,D\,+\,3\,(Q_1^2-Q_0^2)\, 
\end{equation}
which yields the condition
\begin{equation}
\frac{c-D}{24}= \frac{\gamma Q-1}{\gamma^2}\,; 
\end{equation}
cf. \cite{Mazzitelli}. In the case $D=2$, one recovers the known relation $
Q=\gamma + {\gamma}^{-1}$ for quantum Liouville field theory.
\begin{figure}[ht]
\centering
\includegraphics[scale=0.52]{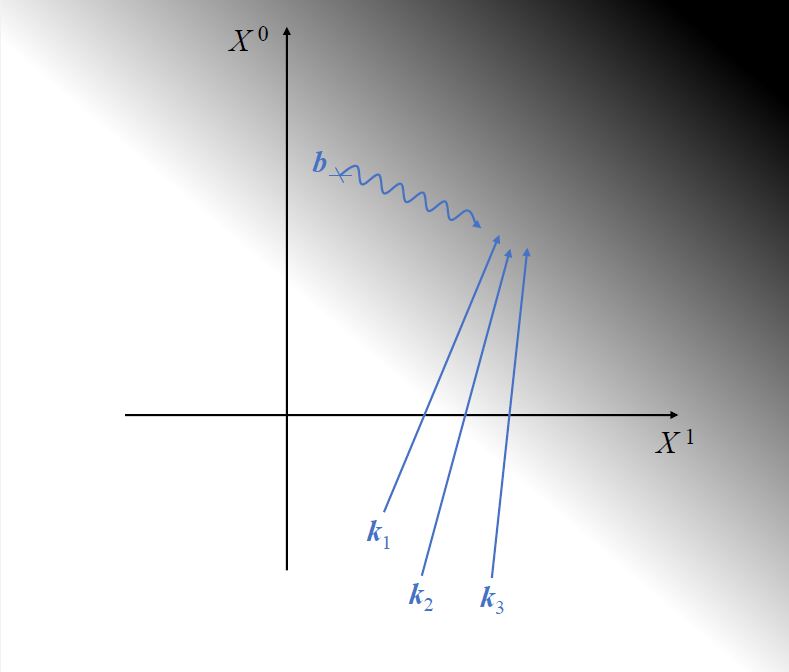}
\caption{Scheme of the Liouville wall disposed along a light-like direction.} \label{Fig1}
\end{figure}

\section{Neutralizing the Liouville mode}

The action of the string $\sigma$-model includes a Liouville contribution
\begin{equation}
S\,\supset \frac{1}{4\pi }\int_Md^2z\,(\,\partial X^-\bar\partial X^++4\pi \mu\, e^{2bX^+}\,)\label{KKKKK}
\end{equation}
along the light-like direction $X^+$, with background charge $Q_-$; we denote $X^{\pm}=X^0\pm X^1$ and $Q_{\pm}=\frac 12 (Q_0\mp Q_1)$.

The fact that the Liouville wall is normal to a light-like direction produces a neutralization of the Liouville mode. This neutralization effect is due to the constant curvature constraint (\ref{vcc}) and was early studied in \cite{Chamseddine, Gaston}. In particular, it makes the tachyonic constituents of the Liouville wall not interact with themselves. More concretely, the screening operators coming from the self-interaction term of Liouville action (\ref{KKKKK}) only see the transverse $X^-$-dependent excitations of the vertex operators involved in the amplitudes, but there are no screening-screening interactions. This is related to the fact that, in this scenario, the condition for the Liouville exponential interaction to be a marginal operator turns out to be linear in $b$, in contrast with the usual quadratic condition \cite{Gaston}. This results in a remarkable simplification of the correlators which, in particular, will allow us to explicitly compute the four-point function.

Before concluding this section, let us mention that the neutralization of the Liouville mode ceases as soon as we introduce additional terms in the JT action that relax the constant-curvature constraint (\ref{vcc}); for instance, if we add
\begin{equation}
S\,\supset \frac{1}{2\pi }\int_Md^2x\sqrt{\hat g}\,(\,\hat g^{\alpha\beta }\, \partial_\alpha \varphi \partial_\beta \varphi \,+\,W(\varphi)\,)\, .
\end{equation}
The introduction of the kinetic term for $\varphi $ would make the Liouville potential to acquire a dependence on $X^-$ and consequently self-interact, while the potential $W(\varphi )$ would introduce new $X^-$-dependent interaction operators in the correlators that would also interact with the Liouville mode.

\section{Correlation functions}

Now, let us move to compute the correlation functions. More specifically, we will compute correlation functions of vertex operators of the form 
\begin{equation}
\Phi_{{k}_j}(z_j)\, = \, e^{2ik_{\mu}^jX^{\mu}(z_j)}\, ,\label{L19}
\end{equation}
with $j=1,2,...,N$ and $\mu=0, 1, ..., D-1$. These are operators that create primary states of conformal weight $k^j_{\mu}k^{\mu\, j}+ik^{j\, \mu}Q_{\mu }$ with momentum $k^j_{\mu}$. We find it convenient to define momenta $ k_{\pm}^j =\frac 12 (k_0^j\mp k_1^j)$, which may take non-real values for normalizable operators in the presence of the background charge.

The $N$-point correlation functions of operators (\ref{L19}) are defined as follows
\begin{equation}
\Big\langle \prod_{j=1}^{N}\Phi_{k_j}(z_j)\Big\rangle =\int\mathcal{D}X\, e^{-S}\, \prod_{j=1}^{N}e^{2ik_{\mu_j}^jX^{\mu_j}(z_j)}\label{vardo}
\end{equation}
where we are omitting ghost dependence, radial ordering symbols and other decorations. Integrating over the zero-mode of the Liouville field \cite{GL}, these $N$-point correlation functions can be expressed as follows
\begin{eqnarray}
\Big\langle \prod_{j=1}^{N}\Phi_{k_j}(z_j)\Big\rangle &=&\frac{\Gamma(-n)\mu^n}{b}e^{-2\theta}\int \prod_{r=1}^{n}d^2w_r\,\int\mathcal{D}X\, e^{-S_{0}}\, \nonumber\\
&& \times \,\prod_{j=1}^{N} e^{2ik_{\mu_j}^j X^{\mu_j}(z_j)}
\prod_{r=1}^{n}e^{2bX^+(w_r)}\, \nonumber \\
&& \times \,\delta\left( \sum_{j=1}^Nk_+^j+iQ_+ \right)\,
\delta\left(\sum_{j=1}^Nk_a^j \right)\,\nonumber
\end{eqnarray}
where $a=2,3,...,D-1$ and $n$ is given by
\begin{equation}
n\,b\,+\,i\sum_{j=1}^Nk_-^j=Q_-\, .
\end{equation}

The advantage of the expression above is that it reduces the $N$-point function (\ref{vardo}) to an $(N+n)$-point function of a free theory. Consequently, it can be solved by considering the free field correlator $
\langle X^{\mu}(z_i)X^{\nu}(z_j)\rangle = -\,\eta ^{\mu\nu}\log |z_i-z_j|$,
which in particular yields $\langle X^{-}(z_j)X^{+}(w) \rangle = 2\log |z_i-w|$. In the following subsections, we will explicitly compute the three- and four-point functions using this Coulomb gas method.

\subsection{The tree-point function}

Let us review the computation of the three-point function. We will focus on the so-called ``resonant correlators'', which correspond to $n\in \mathbb{Z}_{>0}$. Such a three-point function can be computed using the free field propagators given above; it yields \cite{Gaston}
\begin{eqnarray}
&&\Big\langle \prod_{j=1}^{3}\Phi_{k_j}(z_j)\Big\rangle_{\text{R}} = \frac{(-\pi \mu )^n}{b\,\Gamma(n+1)}\,e^{-2\theta } \,\Big({I_3}\Big)^n\nonumber \\
&& \ \ \ \ \ \ \ \  \ \ \ \ \ \ \ \ \times\, \, \delta \Big(\sum_{j=1}^3 k_+^j+\frac{i}{2b}\Big)\,\delta \Big(\sum_{j=1}^3 k_a^j\Big)
\label{elR}
\end{eqnarray}
with the conformal integral
\begin{equation}
I_3=\frac{1}{\pi } \int_\mathbb{C} d^2w\, |w|^{8ibk_+^1}|w-1|^{8ibk_+^2}\, ,\label{elVira}
\end{equation}
where we have set $z_1=0$, $z_2=1$, $z_3=\infty $ using $PSL(2,\mathbb{C})$ invariance. Restoring the dependence on $z_j$ in the three-point function is trivial, as fully determined by conformal invariance. The subscript $\text{R}$ on the left-hand side of (\ref{elR}) stands for ``residue'': resonant correlators sit on poles, so we have used that
\begin{equation}
\lim _{\varepsilon \to 0} \frac{\Gamma (\varepsilon-n)}{\Gamma (\varepsilon)} = \frac{(-1)^n}{\Gamma (n+1)}.
\end{equation}
The $\delta$-functions in (\ref{elR}) come from the integration of the zero modes of the fields. The correlator does not vanish provided $i(k_-^1+k_-^2+k_-^3)=Q_--nb$, $k_+^1+k_+^2+k_+^3=-iQ_+$ and $k_a^1+k_a^2+k_a^3=0$ for $a=2,3,...,D-1$.

Expression (\ref{elR})-(\ref{elVira}) represents a remarkable simplification with respect to the standard Liouville three-point function, which typically involves a Dotsenko-Fateev multiple integral, cf. eq. (B.9) in ref. \cite{DF} and eq. (7.4.2) in ref. \cite{Dotsenko}. However, the neutralization of the Liouville field allows the integrals to factorize and reduce to a product of $n$ Shapiro-Virasoro integrals (\ref{elVira}), which are well-known by any string theorist. This yields
\begin{eqnarray}
&&\Big\langle \prod_{j=1}^{3}\Phi_{k_j}(z_j)\Big\rangle_{\text{R}} = 
\frac{(-\pi \mu )^n}{b\,\Gamma (n+1)}\,e^{-2\theta } \,\prod_{j=1}^3 \frac{\Gamma^n(1+4ibk_+^j)}{\Gamma^n(-4ibk_+^j)}\, \nonumber \\
&& \ \ \ \ \ \ \ \ \ \ \ \ \ \ \ \ \ \ \ \ \times\, \, \delta \Big(\sum_{j=1}^3 k_+^j+\frac{i}{2b}\Big)\,\delta \Big(\sum_{j=1}^3 k_a^j\Big),\label{3pp}
\end{eqnarray}
with $nb+i(k_-^1+k_-^2+k_-^3)=Q_-$. This expression exhibits poles of order $n$ at $k_+^j= {ib\,n}/{4}$ with $n\in \mathbb{Z}_{>0}$, and zeros for $n\in \mathbb{Z}_{\leq 0}$. This three-point function has been computed in \cite{Gaston}. We will see below how the same techniques can be applied to obtain the four-point function.

\subsection{The four-point function}

The four-point function takes the form
\begin{eqnarray}
&&\Big\langle \prod_{j=1}^{4}\Phi_{k_j}(z_j)\Big\rangle_{\text{R}} = \frac{(-\pi\mu)^n}{b\,\Gamma(n+1)}\,e^{-2\theta } \,|z|^{4k^1_{\mu}k^{2\,\mu}}|1-z|^{4k^2_{\mu}k^{3\,\mu}}\, \,\nonumber \\
&& \ \ \ \ \ \ \ \ \ \ \times\, \,\Big(I_4(z)\Big)^n\,  \delta \Big(\sum_{j=1}^4 k_+^j+\frac{i}{2b}\Big) \delta \Big(\sum_{j=1}^4 k_a^j\Big)
\label{elRR}
\end{eqnarray}
where now the conformal integral is
\begin{equation}
I_4(z)=\frac{1}{\pi } \int_\mathbb{C} d^2w\, |w|^{8ibk_+^1}|w-z|^{8ibk_+^2}|w-1|^{8ibk_+^3}\,\label{pizzeria} 
\end{equation}
We have $i(k_-^1+k_-^2+k_-^3+k_-^4)=Q_--nb$ for $n\in \mathbb{Z}_{>0}$, $k_+^1+k_+^2+k_+^3+k_+^4=-iQ_+$, and $k_a^1+k_a^2+k_a^3+k_a^4=0$ with $a=2,3,...,D-1$. Now we have set $z_1=0$, $z_2=z$, $z_3=1$, $z_{4}=\infty $; $z$ is the cross ratio. Integral $I_4$ depends on $z$ and turns out to be not as simple as the Shapiro-Virasoro integral we encountered in the calculation of the three-point function. Despite this, it can still be computed explicitly in terms of hypergeometric functions. This yields
\begin{eqnarray}
&&    {I}_4 (z)     = C_1 (k_+^j )\, \left|\pFq{2}{1}{-4ib k_+^2,1+4ib k_+^4}{-4ib (k_+^1 + k_+^2)}{z}\right|^2\nonumber \\
&&    \  + \, C_2(k_+^j )\,\left|z^{1+4ib (k_+^1 + k_+^2)}\pFq{2}{1}{-4ib k_+^3,1+4ib k_+^1}{-4ib (k_+^3+k_+^4)}{z}\right|^2\nonumber
\end{eqnarray}
with the coefficients
\begin{eqnarray}
C_1(k_+^j )&=&\frac{\Gamma(1+4ib (k_+^1 + k_+^2)) \Gamma(1 + 4ib k_+^3) \Gamma(1+4ib k_+^4) }{\Gamma(-4ib (k_+^1 + k_+^2)) \Gamma(-4ib k_+^3) \Gamma(-4ib k_+^4) }\nonumber \\
    C_2(k_+^j )&=& \frac{\Gamma(1+4ib (k_+^3+k_+^4))\Gamma(1+4ib k_+^1)\Gamma(1+4ib k_+^2))}{\Gamma(-4ib (k_+^3+k_+^4))\Gamma(-4ib k_+^1)\Gamma(-4ib k_+^2)}\nonumber
    \\\nonumber
\end{eqnarray}
see eq. (7.4.1) and eqs. (7.4.19)-(7.4.22) of ref. \cite{Dotsenko}. We used $k_+^1+k_+^2+k_+^3+k_+^4=-i/2b$ to include $k_+^4$ in the above expressions. 

\subsection{Comments}

Expression (\ref{elRR})-(\ref{pizzeria}) is the four-point resonant correlation function of the theory. Let us analyze some properties of it:

{\it Monodromy invariance} of the four-point function (\ref{elRR})-(\ref{pizzeria}) can easily be proven by noticing that the ratio
\begin{equation}
\lambda =\frac{C_2(k_+^j)}{C_1(k_+^j)}
\end{equation}
precisely coincides with the relative coefficient of the independent solutions of the hypergeometric equation that makes the solution invariant; see, for instance, eqs. (4.20)-(4.21) in ref. \cite{MO3}. That suffices to prove the monodromy invariance of the resonant correlators. 

{\it Crossing symmetry} and the braiding coefficients can be analyzed using the Kummer relations of the hypergeometric function. Firstly, notice that, under the crossing operation $k_+^{1,2}\leftrightarrow k_+^{3,4}$, the coefficients above change as $C_1(k_+^j)\leftrightarrow C_2(k_+^j)$. Secondly, the Euler relation of the hypergeometric function simply realizes the symmetry under $k_+^1\leftrightarrow k_+^2$ leaving $z$ invariant. Finally, we observe that similar relations for the hypergeometric function imply that the change $k_+^1\to k_+^3$ in $I_4(z)$ actually implements the transformation $z\to 1-z$, as of course expected.

{\it The three-point function} is recovered from the expression of the four-point function by taking $k_+^2\to 0$. In that limit, the $\Gamma(-4ibk_+^2)$ in the denominator of $C_2(k_+^j)$ makes the second term of $I_4(z)$ vanish, while the hypergeometric function in the first term of $I_4(z)$ tends to 1. Then, the coefficient $C_1(k_+^j)$ exactly reproduces the three-point function (\ref{3pp}); namely
\begin{equation}
\lim_{k_+^2\to 0} I_4(z)=\frac{\Gamma(1+4ibk_+^1)\Gamma(1+4ibk_+^3)\Gamma(1+4ibk_+^4)}{\Gamma(-4ibk_+^1)\Gamma(-4ibk_+^3)\Gamma(-4ibk_+^4)}\, .
\end{equation}

{\it Factorization} properties of the four-point resonant correlation function can also be directly studied using the expansion of the hypergeometric function around $z\simeq 0$.

\section{Final remarks}

Thinking of the JG gravity coupled to Liouville field theory as a non-critical string \cite{Mazzitelli}, we have been able to compute the four-point resonant correlation function in the theory, giving an analytic expression of it in terms of hypergeometric functions. The constant curvature constraint of the JT results in the neutralization of the Liouville mode \cite{Chamseddine}, which ultimately makes the conformal integrals factorize. This made it possible to compute the correlators in a simple way \cite{Gaston}. The four-point function is given in (\ref{elRR})-(\ref{pizzeria}).

We would like to conclude with some open questions: Firstly, it would be interesting to investigate whether a direct connection exists between the model we have considered here and the model studied in \cite{Verlinde1}, where the authors propose a gravity dual to the double scaled Sachdev-Ye-Kitaev (SYK) model consisting of two Liouville theories with complex individual central charges, cf. \cite{Verlinde2}. The holographic duality between JT and SKY suggests that a connection likely exists, although the connection is not clear to us. Secondly, there is another interesting paper that appeared recently \cite{Bruno}, in which light-like Liouville-type potentials are considered within the context of string $\sigma $-models. It would also be interesting to investigate the connection to the results therein. Thirdly, the computation of correlation functions in the case $\Lambda\neq 0$ is also a problem that remains to be explored. The inclusion of a puncture operator in the action makes the Coulomb gas computation of the correlation functions much more involved; however, perturbative techniques such as the one studied in \cite{Zamo} may be of help. Finally, JT gravity has been studied in connection to Liouville field theory in \cite{Liouville1}-\cite{Liouville5}; studying the relation with those recent works would be interesting. 

\[\]
\subsection*{Acknowledgments}
The authors thank Bin Zhu and Tomasz Taylor for correspondence.

\[ \]

\providecommand{\href}[2]{#2}\begingroup\raggedright\endgroup
\end{document}